\documentclass[twocolumn,showpacs,preprintnumbers,amsmath,amssymb,superscriptaddress]{revtex4}

\usepackage{graphicx}
\graphicspath{{fig/}}
\usepackage{dcolumn}
\usepackage{bm}

\begin{document}

\preprint{}

\title{First Results from the XENON10 Dark Matter  Experiment \\at the Gran Sasso
National Laboratory}
\author{J.~Angle}
\affiliation{%
Department of Physics, University of Florida, Gainesville, FL 32611, USA}%
\affiliation{%
Department of Physics, RWTH Aachen University, Aachen, 52074, Germany}

\author{E.~Aprile}%
\email[XENON Spokesperson (E. Aprile). Phone: +1-212-854-3258. \\   E-mail address: ]{age@astro.columbia.edu.}
\affiliation{%
Department of Physics, Columbia University, New York, NY 10027, USA}%

\author{F.~Arneodo}
\affiliation{%
INFN - Laboratori Nazionali del Gran Sasso, Assergi, 67010, Italy}

\author{L.~Baudis}
\affiliation{%
Department of Physics, RWTH Aachen University, Aachen, 52074, Germany}

\author{A.~Bernstein}
\affiliation{%
Lawrence Livermore National Laboratory, Livermore, CA 94550, USA}

\author{A.~Bolozdynya}
\affiliation{%
Department of Physics, Case Western Reserve University, Cleveland, OH 44106, USA}

\author{P.~Brusov}
\affiliation{%
Department of Physics, Case Western Reserve University, Cleveland, OH 44106, USA}

\author{L.C.C.~Coelho}
\affiliation{%
Department of Physics, University of Coimbra, R. Larga, 3004-516, Coimbra, Portugal}

\author{C.E.~Dahl}
\affiliation{%
Department of Physics, Case Western Reserve University, Cleveland, OH 44106, USA}
\affiliation{%
Department of Physics, Princeton University, Princeton, NJ 08540, USA}

\author{L.~DeViveiros}
\affiliation{%
Department of Physics, Brown University, Providence, RI 02912, USA}

\author{A.D.~Ferella}
\affiliation{%
Department of Physics, RWTH Aachen University, Aachen, 52074, Germany}
\affiliation{%
INFN - Laboratori Nazionali del Gran Sasso, Assergi, 67010, Italy}

\author{L.M.P.~Fernandes}
\affiliation{%
Department of Physics, University of Coimbra, R. Larga, 3004-516, Coimbra, Portugal}

\author{S.~Fiorucci}
\affiliation{%
Department of Physics, Brown University, Providence, RI 02912, USA}

\author{R.J.~Gaitskell}
\affiliation{%
Department of Physics, Brown University, Providence, RI 02912, USA}

\author{K.L.~Giboni}
\affiliation{%
Department of Physics, Columbia University, New York, NY 10027, USA}%

\author{R.~Gomez}
\affiliation{%
Department of Physics, Rice University, Houston, TX, 77251, USA}

\author{R.~Hasty}
\affiliation{%
Department of Physics, Yale University, New Haven, CT 06511, USA}

\author{L.~Kastens}
\affiliation{%
Department of Physics, Yale University, New Haven, CT 06511, USA}

\author{J.~Kwong}
\affiliation{%
Department of Physics, Case Western Reserve University, Cleveland, OH 44106, USA}
\affiliation{%
Department of Physics, Princeton University, Princeton, NJ 08540, USA}

\author{J.A.M.~Lopes}
\affiliation{%
Department of Physics, University of Coimbra, R. Larga, 3004-516, Coimbra, Portugal}

\author{N.~Madden}
\affiliation{%
Lawrence Livermore National Laboratory, Livermore, CA 94550, USA}

\author{A.~Manalaysay}
\affiliation{%
Department of Physics, University of Florida, Gainesville, FL 32611, USA}%
\affiliation{%
Department of Physics, RWTH Aachen University, Aachen, 52074, Germany}

\author{A.~Manzur}
\affiliation{%
Department of Physics, Yale University, New Haven, CT 06511, USA}

\author{D.N.~McKinsey}
\affiliation{%
Department of Physics, Yale University, New Haven, CT 06511, USA}

\author{M.E.~Monzani}
\affiliation{%
Department of Physics, Columbia University, New York, NY 10027, USA}%

\author{K.~Ni}
\affiliation{%
Department of Physics, Yale University, New Haven, CT 06511, USA}

\author{U.~Oberlack}
\affiliation{%
Department of Physics, Rice University, Houston, TX, 77251, USA}

\author{J.~Orboeck}
\affiliation{%
Department of Physics, RWTH Aachen University, Aachen, 52074, Germany}

\author{G.~Plante}
\affiliation{%
Department of Physics, Columbia University, New York, NY 10027, USA}%

\author{R.~Santorelli}
\affiliation{%
Department of Physics, Columbia University, New York, NY 10027, USA}%

\author{J.M.F.~dos~Santos}
\affiliation{%
Department of Physics, University of Coimbra, R. Larga, 3004-516, Coimbra, Portugal}

\author{P.~Shagin}
\affiliation{%
Department of Physics, Rice University, Houston, TX, 77251, USA}

\author{T.~Shutt}
\affiliation{%
Department of Physics, Case Western Reserve University, Cleveland, OH 44106, USA}

\author{P.~Sorensen}
\affiliation{%
Department of Physics, Brown University, Providence, RI 02912, USA}

\author{S.~Schulte}
\affiliation{%
Department of Physics, RWTH Aachen University, Aachen, 52074, Germany}

\author{C.~Winant}
\affiliation{%
Lawrence Livermore National Laboratory, Livermore, CA 94550, USA}

\author{M.~Yamashita}
\affiliation{%
Department of Physics, Columbia University, New York, NY 10027, USA}%

\collaboration{XENON Collaboration}

\date{\today}
\begin{abstract}
The XENON10 experiment at the Gran Sasso National Laboratory uses a 15 kg xenon dual phase time projection chamber (XeTPC) to search for dark matter  weakly interacting massive particles (WIMPs).  The detector measures simultaneously the scintillation and the ionization produced by radiation in pure liquid xenon, to discriminate signal from background down to 4.5 keV nuclear recoil energy. A blind analysis of 58.6 live days of data, acquired between October 6, 2006 and February 14, 2007, and using a fiducial mass of 5.4 kg, excludes previously unexplored parameter space, setting a new 90\% C.L. upper limit for the WIMP-nucleon  spin-independent cross-section of $\rm 8.8 \times 10^{-44}\,cm^{2}$ for a WIMP mass of 100~GeV/$c^{2}$, and $\rm 4.5 \times 10^{-44}\,cm^{2}$ for a WIMP mass of 30~GeV/$c^{2}$. This result further constrains predictions of supersymmetric models.
\end{abstract} 

\maketitle


The well-established evidence for non-baryonic dark matter \cite{CMB, SDSS, nuclsynth}  is a striking motivation for physics beyond the Standard Model of particle physics. Weakly interacting massive particles  (WIMPs)~\cite{WIMPs} as  dark matter candidates arise naturally in various theories, such as Supersymmetry, Extra Dimensions, and Little Higgs models~\cite{Bottino,Ellis,ExtraDimns,LittleHiggs}. Since by hypothesis the WIMPs interact through the weak interaction and can efficiently transfer kinetic energy by elastically scattering from atomic nuclei, the WIMP model can be tested by searching for nuclear recoils in a sensitive, low-radioactivity detector~\cite{Jun96,WIMPdet}.  Predicted event rates are less than 0.1~events/kg/day, with energy depositions of the order of 10~keV.

XENON10 is the first 3-D position sensitive TPC developed within the XENON program to search for dark matter WIMPs in liquid xenon (LXe) \cite{XENON}.  Dual phase operation enables the simultaneous measurement of  direct scintillation in the liquid and of ionization, via proportional scintillation in the gas \cite{Bolozdynya}. The ratio of the two signals is different for nuclear (from WIMPs and neutrons)  and electron (from gamma and beta background) recoil events~\cite{Yamashita:2003}, providing event-by-event discrimination down to a few keV nuclear recoil energy \cite{Aprile:2006kx, Shutt:2007}.  In March 2006 the detector was deployed underground at the  Gran Sasso National Laboratory (LNGS) \cite{LNGS}, where it has been in continuous operation for a period of about 10~months, with excellent stability and performance \cite{Xe10_Instr}. The TPC active volume is defined by  a Teflon cylinder of 20 cm inner diameter and  15 cm  height. Teflon is used as an effective UV light reflector \cite{Yamashita:04} and  electrical insulator.  Four stainless steel (SS) mesh electrodes, two in the liquid and two in the gas, with appropriate bias voltages, define the electric fields to drift ionization electrons in the liquid, extract them from the liquid surface and accelerate them in the gas gap. For the dark matter search reported here, the drift field in the liquid was 0.73~kV/cm.   

Two arrays of 2.5~cm square, compact metal-channel photomultiplier tubes (PMTs) (Hamamatsu R8520-06-Al) detect both the direct ($S1$) and proportional ($S2$) scintillation light. The bottom array of 41 PMTs is in the liquid, 1.5 cm below the cathode mesh, to efficiently collect the majority of the direct light which is preferentially reflected downwards at the liquid-gas interface. The top array of 48 PMTs, in the gas, detects the majority of the proportional scintillation light. From the distribution of the PMT hits on the top array, the event location in $XY$ can be reconstructed with a position resolution of a few millimeters. The third coordinate is inferred from the electron drift time measured across 15 cm of LXe, with better than 1 millimeter resolution. The PMTs are digitized at 105 MHz with the trigger provided by the $S2$ sum of 34 center PMTs of the top array. 

The TPC is enclosed in a SS  vessel, insulated by a vacuum cryostat, also made of SS. Reliable and stable cryogenics is provided by a pulse tube refrigerator (PTR)~\cite{Haruyama:2004} with sufficient  cooling power to liquefy the Xe gas and maintain the liquid temperature at -93~$\rm ^o$C. The Xe gas used for the XENON10 experiment was commercially procured with a guaranteed Kr level below 10 part per billion (ppb).  The XENON10 detector is surrounded by a shield made of 20 cm-thick polyethylene and 20 cm-thick lead, to reduce background from external  neutrons and gamma-rays. At the Gran Sasso depth of 3100 meters water equivalent, the surface muon flux is reduced by a factor of 10$^{6}$, such that a muon veto was not necessary for the sensitivity reach of XENON10. 

We report here the analysis of 58.6 live-days of WIMP-search data taken with the XENON10 detector at  Gran Sasso during the period between October 6, 2006 and February 14, 2007. The analysis was performed ``blind", i.e. the events in and near the signal region were not analyzed until the final signal acceptance window and event cuts were tested and defined, using low energy electron and nuclear recoils from calibration data, as well as 40  live-days of  ``unmasked" WIMP-search data. In this letter we interpret the data in terms of spin-independent WIMP-nucleon scattering cross-section. Another letter will focus on the spin-dependent interpretation of the same data~\cite{Yamashita:2007}. More details on the analysis and the estimation of backgrounds will be reported elsewhere \cite{Xe10_Instr}.

A total of $\sim10^{4}$ electron recoil events, in the \textit{a priori} set energy range of interest (4.5 to 26.9~keV nuclear recoil equivalent energy) for the WIMP search, were collected with $^{137}$Cs source. The number of events in the final fiducial volume is about 2400, 1.3 times the statistics of the WIMP-search data. The detector's response to nuclear recoils was obtained from 12~hours irradiation in-situ, using a 200 n/s AmBe source. 

The $S1$ signal associated with each triggered event is searched for in the off-line analysis. By requiring a coincident signal in at least two PMTs, the efficiency of the $S1$ signal search algorithm is larger than 99\%, above a threshold of 4.4 photoelectrons~(pe), or 4.5 keV nuclear recoil equivalent energy. The $S2$ hardware trigger threshold is 100~pe, corresponding to about 4 electrons extracted from the liquid, which is the expected charge from an event with less than 1 keV nuclear recoil equivalent energy \cite{Aprile:2006kx}. The $S2$ trigger efficiency, with a software threshold of 300~pe, is more than 99\% for 4.5~keV nuclear recoils. Basic-quality cuts, tuned on calibration data, are used to remove uninteresting events (e.g. multiple scatter and missing $S2$ events), with a cut acceptance for single-scatter events close to~99\%.

Energy calibration was obtained with  an external $^{57}$Co gamma ray source and with gamma rays from metastable Xe isotopes produced by neutron activation of a 450~g Xe sample, introduced into the detector after the WIMP search data taking. The $S1$ and $S2$ response from the $\rm^{131m}Xe$ 164~keV gamma rays, which interact uniformly within the detector, were used to correct $\pm$20\% variations of the signals due to the position dependence of the light collection efficiency. $S1$ for 122~keV gamma rays, after position-dependent corrections, gives a volume-averaged light yield, $L_y$, of $3.0 \pm 0.1$(syst)$\pm0.1$(stat)~pe/keVee (keVee is the unit for electron-recoil equivalent energy) at the drift field of 0.73~kV/cm. The nuclear recoil equivalent energy (in unit of keV) can be calculated as $E_{nr} = S1/L_y/\mathcal{L}_{eff}\cdot S_e/S_n$. Here $\mathcal{L}_{eff}$ is the nuclear-recoil scintillation efficiency relative to that of 122 keV gamma rays in LXe at zero drift field. We used a constant $\mathcal{L}_{eff}$ value of 0.19 which is a simple assumption consistent with the most recent measurements \cite{Aprile:2005mt, Chepel:2006}. $S_e$ and $S_n$ are the scintillation quenching factors due to the electric field, for electron and nuclear recoils, respectively. $S_e$ and $S_n$ were measured to be 0.54 and 0.93, respectively, at a drift field of 0.73~kV/cm \cite{Aprile:2006kx}.

\begin{figure}
\includegraphics[width=0.42\textwidth]{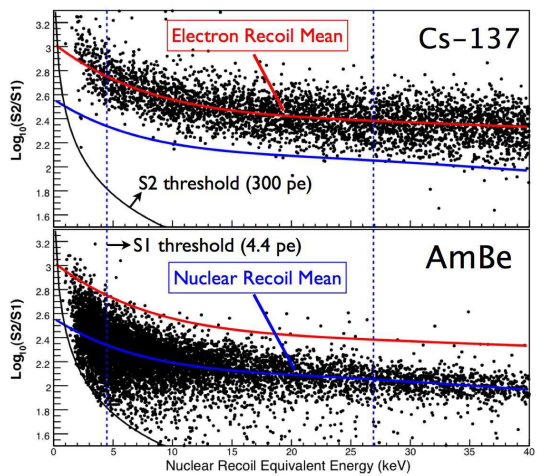}
\caption{\label{fig:band} Log$_{10}$($S2/S1$) as a function of energy for electron recoils (top) and nuclear recoils (bottom) from calibration data. The colored lines are the mean Log$_{10}$($S2/S1$) values of the electron recoil (upper, red) and nuclear recoil (lower, blue) bands. The region between the two vertical dashed lines is the energy window (4.5 - 26.9 keV nuclear recoil equivalent energy) chosen for the WIMP search. An $S2$ software threshold of 300 pe is also imposed (black lines).}
\end{figure}

Background rejection is based on the ionization/scintillation ($S2/S1$) ratio, which is different for nuclear and electron recoils in LXe. Figure~\ref{fig:band} shows the energy dependence of the logarithm of this ratio for electron recoils from $^{137}$Cs gamma ray and for nuclear recoils from AmBe fast neutron calibrations. The separation of the mean Log$_{10}$($S2/S1$) values between electron and nuclear recoils increases at lower energy. In addition, the width of the electron recoil band is also smaller at lower energy. The combination of these two effects gives a better electron recoil rejection efficiency at the lower energy, reaching 99.9\%. The different ionization density and track structure of low energy electrons and Xe ions in LXe result in different recombination rate and associated fluctuations, which might explain the observed behavior.

The  Log$_{10}$($S2/S1$) values increase with decreasing energy for both electron and nuclear recoils, within the energy window of interest (4.5 - 26.9 keV nuclear recoil equivalent energy). The same behavior has been observed previously in small prototype detectors~\cite{Aprile:2006kx, Shutt:2007}. In our analysis, we subtract the energy-dependent mean Log$_{10}$($S2/S1$) from the electron recoil band to obtain  $\Delta$Log$_{10}$($S2/S1$) for all events. After this band flattening, the energy window of interest for the WIMP search is divided into seven individual energy bins (see Table~\ref{tab:table1}). For each energy bin, the nuclear recoil acceptance window is defined to be between $\Delta$Log$_{10}$($S2/S1$)~$ = \mu$ and $\Delta$Log$_{10}$($S2/S1$)~$ = \mu-3\sigma$. Here $\mu$ and $\sigma$ are the mean and sigma from a Gaussian fit of the nuclear recoil $\Delta$Log$_{10}$($S2/S1$) distribution. The nuclear recoil acceptance efficiency is the fraction of nuclear recoil events within the acceptance window. The $\Delta$Log$_{10}$($S2/S1$) distribution for electron recoils from the $^{137}$Cs data is found empirically to be statistically consistent with Gaussian fits, except for a small number of ``anomalous leakage events".  From these fits, we estimate the electron recoil rejection efficiency and predict the number of statistical leakage events in the WIMP search data, for the defined nuclear recoil acceptance window. For each energy bin, the derived electron recoil rejection efficiency and the nuclear recoil acceptance values are listed in Table~\ref{tab:table1}.

\begin{table}
\caption{\label{tab:table1} The software cut acceptance of nuclear recoils $\varepsilon_{c}$, the nuclear recoil acceptance $A_{nr}$, and the electron recoil rejection efficiency $R_{er}$  for each of the seven energy bins ($E_{nr}$ in nuclear recoil equivalent energy). The expected number of leakage events, $N_{leak}$, is based on $R_{er}$ and the number of detected events, $N_{evt}$, in each energy bin, for the 58.6 live-days WIMP-search data, with 5.4 kg fiducial. Errors are the statistical uncertainty from the Gaussian fits on the electron recoil $\Delta$Log$_{10}$($S2/S1$) distribution.}
\begin{ruledtabular}
\begin{tabular}{llllrl}
$E_{nr}$ (keV) & $\varepsilon_{c}$  &   $A_{nr}$  &  1 - $R_{er}$ & $N_{evt}$ & $N_{leak}$ \\
   &   &   &  ($10^{-3}$) & & \\
\hline
4.5 - 6.7 & 0.94 &  0.45 &  0.8$^{+0.7}_{-0.4}$ & 213 & 0.2$^{+0.2}_{-0.1}$\\
6.7 - 9.0 & 0.90 &  0.46 &  1.7$^{+1.6}_{-0.9}$  & 195 & 0.3$^{+0.3}_{-0.2}$ \\
9.0 - 11.2 & 0.89 & 0.46 & 1.1$^{+0.9}_{-0.5}$  & 183 & 0.2$^{+0.2}_{-0.1}$ \\
11.2 - 13.4 & 0.85 & 0.44 & 4.1$^{+3.6}_{-2.0}$ & 190 & 0.8$^{+0.7}_{-0.4}$\\
13.4 - 17.9 & 0.83 & 0.49 & 4.2$^{+1.8}_{-1.3}$ & 332 & 1.4$^{+0.6}_{-0.4}$\\
17.9 - 22.4 & 0.80 & 0.47 & 4.3$^{+1.7}_{-1.2}$  & 328 & 1.4$^{+0.5}_{-0.4}$ \\
22.4 - 26.9 & 0.77 & 0.45 & 7.2$^{+2.4}_{-1.9}$ &  374 & 2.7$^{+0.9}_{-0.7}$\\
\hline
Total &     &    &  & 1815 & 7.0$^{+1.4}_{-1.0}$ \\
\end{tabular}
\end{ruledtabular}
\end{table}

In addition to the statistical events leaking from the electron recoil band into the nuclear recoil acceptance window, we observed anomalous leakage events in the $^{137}$Cs calibration data and unmasked WIMP search data. These events were identified to be multiple-scatter events with one scatter in the non-active LXe mostly below the cathode and a second scatter in the active LXe volume. The $S2$ signal from this type of event is from the interaction in the active volume only, while the $S1$ signal is the sum of the two $S1$'s in both the active and non-active volume. The result is a smaller $S2/S1$ value compared to that for a single-scatter event, making some of these events appear in the WIMP-search window. Two types of cuts, one using the $S1$ signal asymmetry between the top and bottom PMT arrays and the other using the $S1$ hit pattern, defined as $S1_{RMS} = \sqrt{\frac{1}{n}\sum(S1_i - \overline{S1})^2} ~(i = 1, n)$, on either the bottom or the top PMT array, are defined to remove these anomalous events. The $S1$ signal from the scatter outside the active volume tends to be clustered on a few of the bottom PMTs (larger $S1_{RMS}$), while the $S1$ signal from a normal event in the active volume is distributed more evenly over the PMTs (smaller $S1_{RMS}$). A large fraction of events that  leaked into the WIMP-signal window are of this type of background and could be removed by the  cuts discussed above. The cut acceptance $\varepsilon_{c}$ for single-scatter nuclear recoil events, based on AmBe neutron calibration data, is listed in Table~\ref{tab:table1}. 

\begin{figure}[htpb]
\centering
\includegraphics[width=0.4\textwidth]{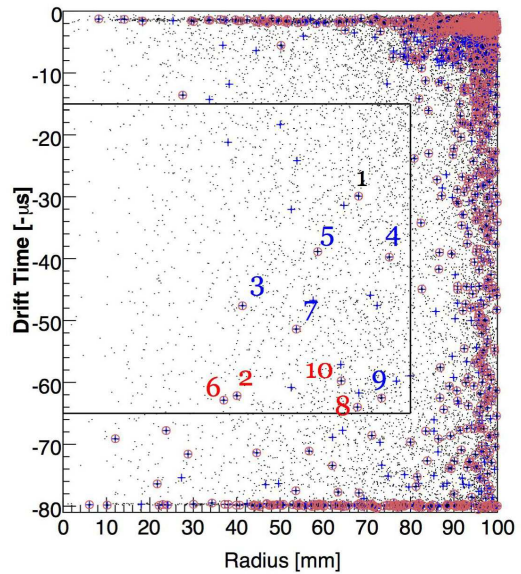}
\caption{
	\label{fd} Position distribution of events in the 4.5 to 26.9~keV nuclear recoil energy window, from the 58.6 live-days of WIMP-search data. ($+$) Events in the WIMP-signal region before the software cuts. ($\oplus$) Events remaining in the WIMP-search region after the software cuts. The solid lines indicate the fiducial volume, corresponding to a mass of 5.4~kg. }
\end{figure}

The 3D position sensitivity of the XENON10 detector gives additional background suppression with fiducial volume cuts \cite{Sorensen}. Due to the high stopping power of LXe, the background rate in the central part of the detector is lower (0.6~events/keVee/kg/day) than that near the edges (3~events/keVee/kg/day). For this analysis, the fiducial volume is chosen to be within 15 to 65~$\mu$s (about 9.3~cm in $Z$, out of the total drift distance of 15~cm) drift time window and with a radius less than 8 cm (out of 10 cm) in $XY$, corresponding to a total mass of 5.4~kg (Fig.~\ref{fd})~\cite{Yamashita:IDM}. The cut in $Z$ also removes many anomalous events due to the LXe around the bottom PMTs, where they happen more frequently compared to the top part of the detector. 

\begin{figure}
\includegraphics[width=0.45\textwidth]{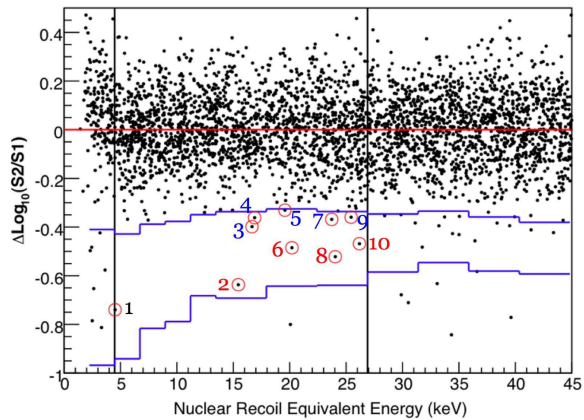}
\caption{\label{fig:wimp} Results from 58.6 live-days of WIMP-search in the 5.4~kg LXe target. The WIMP search window was defined between the two vertical lines (4.5 to 26.9 keV nuclear recoil equivalent energy) and blue lines (about 50\% nuclear recoil acceptance).}
\end{figure}

\begin{figure}
\includegraphics[width=0.45\textwidth]{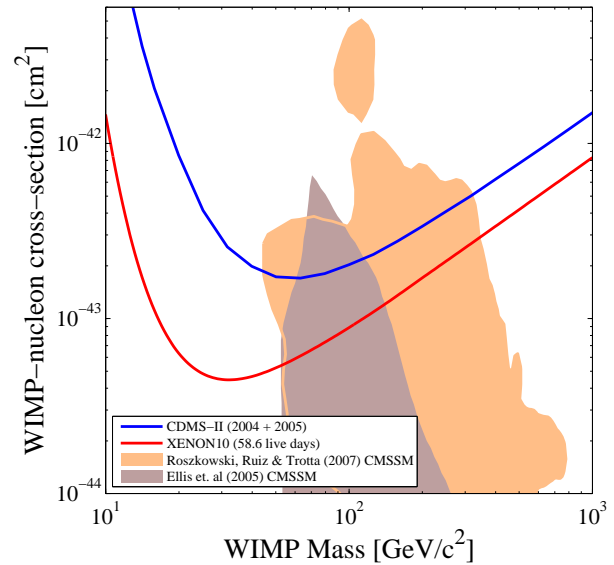}
\caption{\label{fig:limit} Spin-independent WIMP-nucleon cross-section upper limits (90\% C.L.) versus WIMP mass. Curves are shown for the previous best published limit (upper, blue) \cite{CDMSII} and the current work (lower, red), assuming a constant 19\% $\mathcal{L}_{eff}$. The shaded area is for parameters in the constrained minimal supersymmetric models~\cite{Ellis,Roszkowski:2007fd}.}
\end{figure}

After all the cuts were finalized for the energy window of interest, we analyzed the 58.6~live-days of WIMP-search data. From a total of about 1800 events, ten events were observed in the WIMP search window after cuts (Fig.~\ref{fig:wimp}). We expect about seven statistical leakage events (see Table~\ref{tab:table1}) by assuming that the $\Delta$Log$_{10}$($S2/S1$) distribution from electron recoils is purely Gaussian, an assumption which is statistically consistent with the available calibration data, except for a few ``anomalous leakage events". However, the uncertainty of the estimated number of leakage events for each energy bin in the analysis of the WIMP search data is currently limited by available calibration statistics. Based on the analysis of multiple scatter events, no neutron induced recoil event is expected in the single scatter WIMP-search data set. To set conservative limits on WIMP-nucleon spin-independent cross section, we consider all ten observed events, with no background subtraction. Figure~\ref{fig:limit} shows the 90\%~C.L. upper limit on WIMP-nucleon cross sections as a function of WIMP mass, calculated for a constant 19\% $\mathcal{L}_{eff}$, the standard assumptions for the galactic halo \cite{Lewin}, and using the ``maximum gap" method in \cite{Yellin}. For a WIMP mass of 100~GeV/$c^2$,  the upper limit is 8.8~$\times~10^{-44}$~cm$^2$, a factor of 2.3 lower than the previously best published limit \cite{CDMSII}. For a WIMP mass of 30~GeV/$c^2$, the limit is $4.5 \times 10^{-44}$~cm$^{2}$. Energy resolution has been taken into account in the calculation. The largest systematic uncertainty is attributed to the limited knowledge of $\mathcal{L}_{eff}$ at low nuclear recoil energies. Our own measurements of this quantity \cite{Aprile:2005mt} did not extend below 10.8 keVr, yielding a value of ($13.0\pm2.4$)\% at this energy. More recent measurements by Chepel et al. \cite{Chepel:2006} have yielded a value of 34\% at 5 keVr, with a large error.

A comparison between the XENON10 neutron calibration data and Monte Carlo simulated data, including the effects of detector resolution and energy dependence of $\mathcal{L}_{eff}$, provides an effective constraint on the variation of $\mathcal{L}_{eff}$ for all energies in the analysis range~\cite{Angel}.  The constant $\mathcal{L}_{eff}$ assumption used to calculate the limits above shows reasonable agreement at the 10\% level between the Monte Carlo predicted spectrum and the measured energy dependence and intensity of the single scatter nuclear recoil spectrum. The $\mathcal{L}_{eff}$ assumption which gives the best agreement implies a  slightly more sensitive exclusion limit, and is not quoted. A conservative exclusion limit was calculated by including estimates of possible systematic uncertainty in the signal acceptance near threshold. Also included was an estimate of the uncertainty in the energy dependence of the neutron scattering cross sections used in the Monte Carlo simulations. The $\mathcal{L}_{eff}$ assumption which gives poorest sensitivity, while remaining consistent at the 1\% level with the neutron calibration data, corresponds to exclusion limits as high as 10.4~$\times~10^{-44}$~cm$^2$ (5.2~$\times~10^{-44}$~cm$^2$) for a WIMP mass of 100~GeV/$c^2$ (30~GeV/$c^2$).

Although we treated all 10 events as WIMP candidates in calculating the limit, none of the events are likely WIMP interactions. $\Delta$Log$_{10}$($S2/S1$) values for 5 events (compared with 7 predicted) are statistically consistent with the electron recoil band. These are labeled as No.'s 3, 4, 5, 7, 9 in Fig.~\ref{fd} and Fig.~\ref{fig:wimp}. As shown in Table~\ref{tab:table1} these leakage events are more likely to occur at higher energies. {\it A posteriori} inspection of event No.~1 shows that the $S1$ coincidence requirement is met because of a noise glitch.  Event No.'s 2, 6, 8, 10 are not favored as evidence for WIMPs for three main reasons. First, they are all clustered in the lower part of the fiducial volume (see Fig.~\ref{fd}) where anomalous events happen more frequently, as discussed above. Second, the anomalous $S1$ hit pattern cut discussed earlier for the primary blind analysis was designed to be very conservative. An independent secondary blind analysis performed in parallel with the primary analysis, used a more stringent cut to identify anomalous hit patterns in $S1$ and rejected 3 (No.'s 6, 8, 10) of these 4 candidate events. Third, the expected nuclear recoil spectrum for both neutrons and WIMPs falls exponentially with energy, whereas the candidate events appear preferentially at higher energy.

The new XENON10 upper limit on WIMP-nucleon spin-independent cross section  further excludes some parameter space in the minimal supersymmetric models~\cite{Bottino} and the constrained minimal supersymmetric models (CMSSM) (e.g.~\cite{Ellis, Roszkowski:2007fd}). 

This work is supported by the National Science Foundation under grants No.~PHY-03-02646 and PHY-04-00596, and by the Department of Energy under Contract No.~DE-FG02-91ER40688, the CAREER Grant No.~PHY-0542066, the Volkswagen Foundation (Germany) and the FCT Grant No.~POCI/FIS/60534/2004 (Portugal). We thank the Director of the Gran Sasso National Laboratory, Prof. E. Coccia, and his  staff for support throughout this effort. Special thanks go to the Laboratory engineering team, P. Aprili, D. Orlandi and E. Tatananni, and to F.~Redaelli of COMASUD for their contribution to the XENON10 installation. 



\begin{thebibliography}{3}
\bibitem{Yellin} S. Yellin, Phys. Rev. D {\bf 66}, 032005 (2002). The one-sided interval at the 90\% C.L. from this method is about as constraining as a two-sided interval at the 85\% C.L..
\end{thebibliography}

\end{document}